\documentclass[aps,prl,reprint,twocolumn,floatfix,showpacs,superscriptaddress]{revtex4-1}
\pdfoutput=1

\usepackage[author={},hspace=20px,icon=Comment,color=yellow]{pdfcomment}
\usepackage[normalem]{ulem} 
\usepackage{graphicx}
\usepackage{amsmath,amssymb}
\usepackage{natbib}
\usepackage{hyperref}
\usepackage{xcolor}

\usepackage[T1]{fontenc} 

\usepackage[section]{placeins} 

\newcommand{\iu}{{i\mkern1mu}}

\begin{document}
\let\oldcite\cite
\renewcommand{\cite}[1]{\mbox{\oldcite{#1}}}


\title{Transmission of plasmons through a nanowire}

\author{Peter Geisler}
\thanks{equally contributing}
\author{Enno Krauss}
\thanks{equally contributing}
\author{Gary Razinskas}
\thanks{equally contributing}
\affiliation{%
  NanoOptics \& Biophotonics Group, Experimentelle Physik 5, Physikalisches
  Institut, Universit\"at W\"urzburg, Am Hubland, 97074 W\"urzburg, Germany
}
\author{Bert Hecht}
\email[]{hecht@physik.uni-wuerzburg.de}
\affiliation{%
  NanoOptics \& Biophotonics Group, Experimentelle Physik 5, Physikalisches
  Institut, Universit\"at W\"urzburg, Am Hubland, 97074 W\"urzburg, Germany
}
\affiliation{%
  R\"ontgen Research Center for Complex Material Systems (RCCM), Am Hubland,
  97074 W\"urzburg, Germany
}

\date{\today}


\begin{abstract}
Exact quantitative understanding of plasmon propagation along nanowires is
mandatory for designing and creating functional devices. Here we investigate
plasmon transmission through top-down fabricated monocrystalline gold
nanowires on a glass substrate.  We show that the transmission through
finite-length nanowires can be described by Fabry-P\'{e}rot oscillations that
beat with free-space propagating light launched at the incoupling end. Using
an extended Fabry-P\'{e}rot model, experimental and simulated length
dependent transmission signals agree quantitatively with a fully analytical
model.
\end{abstract}


\maketitle



At optical frequencies gold and silver nanowires can be used as subwavelength
waveguides due to the considerably shortened effective wavelength of surface
plasmon polaritons compared to free-space light \cite{novotny_light_1994,
takahara_guiding_1997}.
Nanowires therefore represent fundamental building blocks of nano-optical
circuitry that may find applications in nano quantum optics, information
processing and sensing
\cite{anker_biosensing_2008, martin-cano_resonance_2010,
chang_single-photon_2007, fu_all-optical_2012, raza_electron_2016}.
High-precision experiments of simple physical systems
often reveal subtle but important effects or can be used to test theoretical
descriptions of experimental results.
Validated theoretical descriptions can
then be used with confidence to model more complex systems. Yet, systematic
high-precision experiments of light transmission through nanowires to date
hardly exist. For example, the transmission efficiency of light through such
nanowires in many experiments significantly deviates from theoretical
expectations based on bulk dielectric constants
\cite{kusar_measurement_2012, wild_propagation_2012}.
To date the origin of such
deviations remains unknown since the structural uncertainties of bottom up and
top-down fabricated nanowires are not small enough to allow for conclusive
analyses.

Here we present a systematic study of the monochromatic light transmission
through more than 300 monocrystalline gold nanowires of equal cross section but
variable length ranging between 1940 to 8040~nm in length with a 20~nm
increment.
We demonstrate by experiments and simulations that a quantitatively correct
description of the length-dependent nanowire transmission can be obtained by
also taking into account free-space propagating modes launched by scattering of
the excitation spot at the wire input in addition to Fabry-P\'{e}rot-type internal
plasmon resonances. These free-space propagating modes interact with the
outcoupling end of the wire and beat with the regularly emitted photons
originating from the  wire plasmon's radiative decay. This leads to significant
amplitude modulations of the Fabry-P\'{e}rot transmission resonances. The
quantitative agreement between our model, numerical simulations, and
measurements validates our model and yields values of propagation parameters
that are compatible with bulk dielectric constants and for which remaining
sources of uncertainties are clearly identified.

%
%
\begin{figure}
  \begin{center}
    \includegraphics[width=0.99\columnwidth]{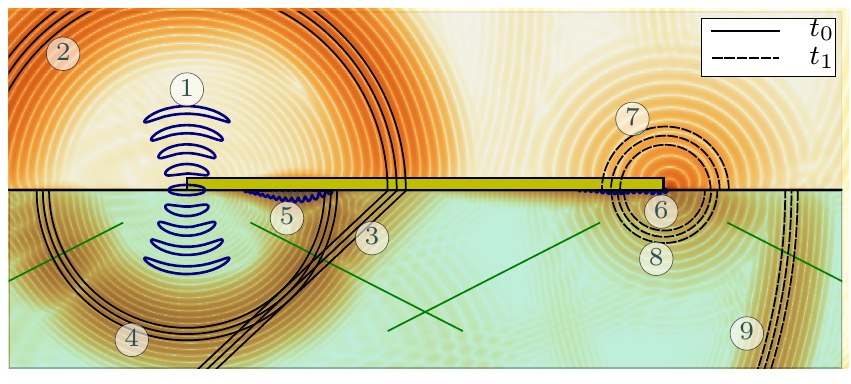}
    \caption{(color online)
      Sketch of the free-space and guided fields involved in the experiment in
      a plane through the nanowire long axis perpendicular to the substrate
      (wire height stretched for better visibility).
      The field intensity distributions are represented in a logarithmic scale
      and were obtained by FDTD simulations. To illustrate progress in time the
      intensities of two short pulses with a short time delay $t_1-t_0$ are
      combined. (1) focused Gaussian source illumination from the glass half
      space; (2) scattered and transmitted light above the glass surface
      leading to (3) refracted waves at the air--glass interface; (4) scattered
      and reflected light below the air--glass interface; (5) launched plasmon
      pulse at time $t_0$; (6) plasmon pulse at time $t_1$; light emission from
      the wire end (7) above and (8) below the air--glass interface; (9)
      scattered and reflected light below the air--glass interface at time
      $t_1$ leading to additional scattered fields at the wire end. The (green)
      lines below the glass surface symbolize the limited accepting angle of
      the objective with NA = 1.3.
    }
    \label{fig:setup}
  \end{center}
\end{figure}
To experimentally characterize the transmission properties of nanowires
of different lengths we use a home-build inverted microscope setup
\cite{geisler_multimode_2013}.
Nanofabricated single crystal gold nanowires supported by a cover glass are
mounted above an oil immersion microscope objective (Leica, 1.30 NA, $\infty$,
PL Fluotar 100x) which is used to focus a laser beam ($\lambda$ = 800~nm, 12~nm
FWHM spectral linewidth, 80~MHz repetition rate, 50~nW average power measured
in front of the objective, NKT Photonics, SuperK Power with SpectraK AOTF) via
a $\lambda/2$-plate (Foctec, AWP210H NIR) to a diffraction-limited (390~nm
diameter) spot at the air--glass interface that is linearly polarized along the
wire axis.
The same objective is used to image the emitted and reflected light onto a CCD
camera (Andor, DV887AC-FI EMCCD) via a 50/50 non-polarizing beamsplitter
(Thorlabs, CM1-BS013). In order to avoid saturation of the CCD, the strong
reflection of the excitation spot is suppressed by a small beam block (OD 2)
introduced in an intermediate image plane.  The exact position of the
excitation spot with respect to the wire end can be adjusted with nm-precision
by moving the sample using a piezo translation stage (Physik Instrumente,
P-527) and was optimized to obtain maximum signal intensity at the wire end.
%
%
The principle of the experiment and all light propagation channels are sketched
in Fig.~\ref{fig:setup} showing free-space and guided fields as obtained from
numerical simulations.  The tightly focused laser at the incoupling end of the
wire (Fig.~\ref{fig:setup}(1)) launches a plasmon with about 30\% efficiency
(Fig.~\ref{fig:setup}(5)) that propagates towards the outcoupling end
(Fig.~\ref{fig:setup}(6)) where it is partly radiated into the surrounding
media (Fig.~\ref{fig:setup}(7) and (8)) while about 40\% of the plasmon field
is reflected and propagates back along the wire leading to Fabry-P\'{e}rot-type
standing waves \cite{fabry_theorie_1899, hofstetter_theory_1997}.
Additional propagating fields - in the following referred to as air-wave and
glass-wave (Fig.~\ref{fig:setup}(2) and (4)) - are launched by the partial
scattering of the excitation source at the incoupling end.
While about 30\% of the laser energy is coupled into the nanowire, about 50\%
of the energy is scattered into the glass half space leading to a spherical
wave originating from the incoupling end of the wire. The remaining 20\% are
transmitted into the air half space where they also evolve as spherical waves.
%
%
\begin{figure}
  \begin{center}
    \includegraphics[width=0.99\columnwidth]{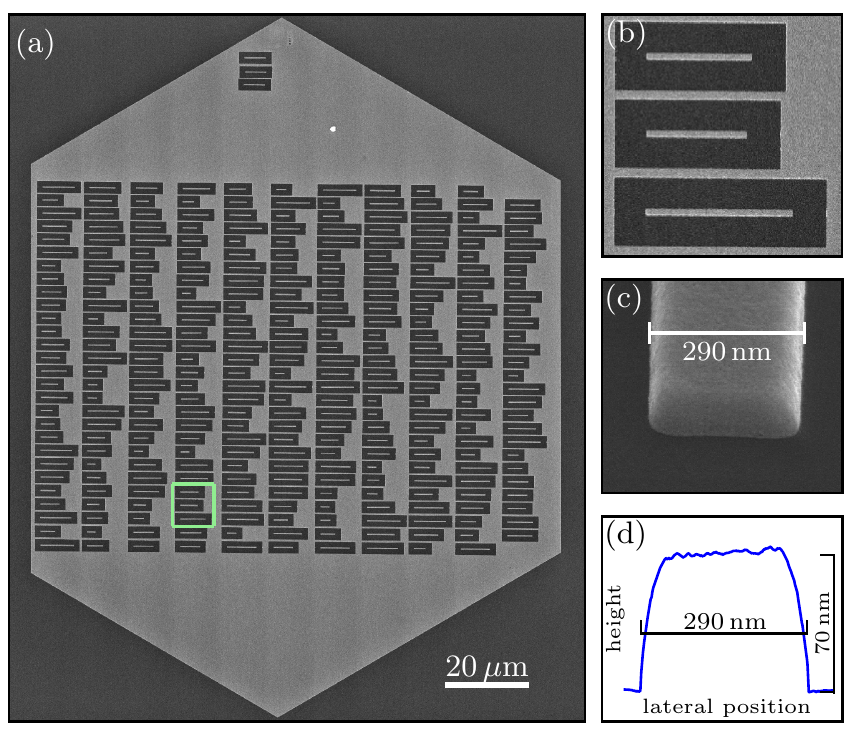}
    \caption{(color online)
      (a)-(c) Scanning electron microscopy (SEM) images of the sample showing
      (a) the full platelet including the focused-ion beam milled area of the
      platelet with the array of single wires of random wire lengths as well as
      (b), (c) two closeups at different zoom levels. (d) Atomic force
      microscopy line profile along one wire's cross section.
    }
    \label{fig:sample}
  \end{center}
\end{figure}
%
%
%
\begin{figure}
  \begin{center}
    \includegraphics[width=0.99\columnwidth]{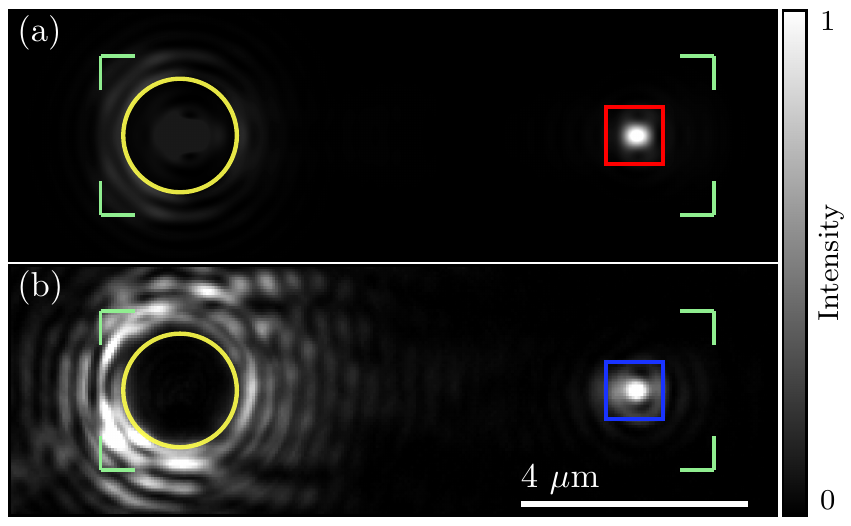}
    \caption{(color online)
      (a) Far-field intensity from FDTD simulation of an 8 $\mu$m long wire and
      (b) the corresponding experimental CCD image. The (yellow) circle
      visualizes the position of the beam block and the (green) corners mark
      the borders between gold platelet on glass and bare glass (see
      Fig.~\ref{fig:sample} and SI). The red (blue) square
      marks the area that was integrated to obtain the intensity in simulation
      (experiment). Experimental data was scaled to match simulated peak
      intensity within the integration area.
    }
    \label{fig:measurement}
  \end{center}
\end{figure}
Another contribution to propagating waves in the glass half space arises from
light refracted at the air--glass interface (Fig.~\ref{fig:setup}(3)) which
propagates into the glass as a plane wave under an angle of about 43$^{\circ}$
(critical angle for total internal reflection) and thus within the acceptance
angle of the objective. This wave leads to a distinct pattern in the wire
far-field images that vanishes for low-NA imaging.
The air-wave and the glass-wave originating from the excitation position are
not directly detected by the camera for different reasons.
While the air-wave propagates away from the collecting objective lens above the
interface the glass-wave is strongly suppressed by the beam block at the
intermediate image plane. However, as we detail below, the interaction of these
propagating waves of different effective wavelengths with the wire end leads to
interference and beating effects in the light intensity emitted
by the wire ends which is detected by the camera.
A video showing the time evolution of a focused laser pulse coupled to the end
of a wire and all resulting free-space and guided fields involved in the experiment
can be found in the SI.

Using an eigenmode solver (Lumerical Solutions, MODE Solutions) it can be confirmed
that the propgation along the present nanowire is single mode at the chosen vacuum wavelength (see SI).
Based on Fabry-P\'{e}rot theory the field amplitude $\psi_{\text{T}}$ that is
transmitted by a waveguide of length $L$ via one single eigenmode and emitted into the 
detection path can be expressed as
\cite{hofstetter_theory_1997}
\begin{equation}
  \psi_{\text{T}} =
  \frac{%
    \psi_{0}~\eta~t~%
    e^{-\left(\alpha+i\beta\right)L}
  }{%
    1-\left(re^{-\left(\alpha+\iu\beta\right)L}\right)^{2}
  }.
  \label{eqn:fpt}
\end{equation}
Here, $\psi_{0}$ is the amplitude of the Gaussian excitation beam, $\eta$ is a
(complex valued) efficiency factor comprising the combined effects of
incoupling into the waveguide and detecting the emitted signal, and $r = R
e^{\iu\phi}$ and $t$ are the complex plasmon reflection and transmission
coefficients, respectively. These coefficients are assumed to be identical for
both wire terminations. Furthermore, $\alpha = 1 / 2 l_{\text{decay}}$ and
$\beta = 2 \pi / \lambda_{\text{eff}}$ denote the propagating wire mode's attenuation and wave
number, respectively, with $l_{\text{decay}}$ being the mode's intensity decay
length and $\lambda_{\text{eff}}$ its effective wavelength.
Both air-wave and glass-wave can be approximated by propagating spherical
waves. The respective amplitudes scattered from the wire far end at a distance
$L$ away from the incoupling end towards the detector can be expressed as
\begin{equation}
  \psi_{\text{medium}} =
  \psi_{0} \eta_{\text{medium}}
  \frac{e^{-\iu \beta_{\text{medium}} L}}{L},
  \label{eqn:beating}
\end{equation}
where $\beta_{\text{medium}} = 2 \pi / \lambda_{\text{medium}}$ is the wave
vector and $\lambda_{\text{medium}}$ is the wavelength of light in the
respective medium, i.e.~air and glass. The complex quantity
$\eta_{\text{medium}}$ denotes a combined efficiency factor accounting for the
efficiency of scattering of the excitation field $\psi_{0}$ at the incoupling
wire end, thus generating the spherical wave in the respective medium, as well
as for the efficiency for scattering of this wave at the wire's far end into
the detection path.
All fields originating from the waveguide end interfere at the detector
according to
\begin{equation}
  I_{\text{total}} =
  | \psi_{\text{T}} + \psi_{\text{air}} + \psi_{\text{glass}} |^{2}.
  \label{eqn:intensity}
\end{equation}

%
%
\begin{figure}
  \begin{center}
    \includegraphics[width=0.99\columnwidth]{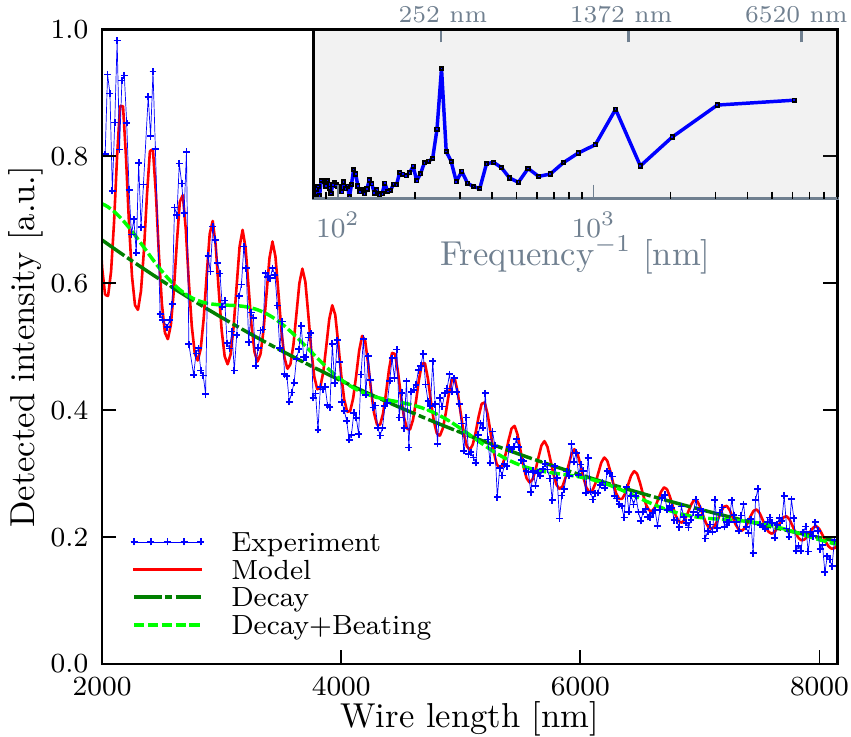}%
    \caption{(color online)
      Detected signal (blue ``x'', thin blue line to guide the eye)
      within the integration box at the wire end.
      Superimposed is the fitted model (solid red line) together with the
      intensity decay curve (interleaved dark green line) and a curve of the
      model with Fabry-P\'{e}rot reflectivity $R=0$ visualizing the
      oscillations from the beating between the Fabry-P\'{e}rot model and the
      scattered free-space waves in air and glass at the wire end (dashed light
      green line);
      (inlay) Fourier transformed experimental data.  The expected peak
      positions from the Fabry-P\'{e}rot oscillation (252~nm), the beating
      between the transmitted light and the air-wave (1372~nm), and the beating
      between the transmitted light and the glass-wave (6520~nm) are indicated.
    }
    \label{fig:impedance}
  \end{center}
\end{figure}
%

%
%
To perform a high-precision experiment that can reveal the effects of the
additional free-space waves on the overall apparent wire transmission we
prepared a sample consisting of 306
monocrystalline \cite{huang_atomically_2010, wu_single-crystalline_2015} gold
nanowires ranging from 1940 nm to 8040 nm in nominal length with a length
increment of 20 nm (Fig.~\ref{fig:sample}). All nanowires were fabricated by
focused-ion beam (FIB) milling of a single monocrystalline gold platelet to
ensure uniform milling conditions. The wire lengths were distributed randomly
over the array to avoid artifacts due to changes of fabrication or measurement
conditions. Simulations show that the remaining gold frames
(Fig.~\ref{fig:sample} (b)) around the wires do not affect the in- or
outcoupled intensity (see SI).
All resulting wires are of uniform quality showing no observable differences
in SEM images apart from the wire length.
To provide optimal conditions for focused-ion beam milling the sample was fabricated
on a conductive substrate (silicon) and then transferred to a clean and flat glass
substrate (no adhesion layers). Transfer to the glass substrate avoids the
presence of a glass ridge below the nanowires and excludes fabrication-induced
surface roughness as well as Ga$^{+}$-Ion implantation within the milling area
\cite{wu_silicagold_2015}.
%
%
The simulation results are obtained by modeling the structure using a
finite-difference time-domain (FDTD) solver (Lumerical Solutions, FDTD Solutions).
The geometry was chosen to match high resolution SEM and AFM images and
includes e.g.\ the soft edges (Fig.~\ref{fig:sample} (c,d)). Optical properties
of the gold are taken from literature \cite{etchegoin_analytic_2006,
etchegoin_erratum_2007}, while the refractive index of the glass was set to $n$
= 1.46. Using the eigenmode solver the
fundamental mode properties for the present geometry at vacuum wavelength
$\lambda$ = 800 nm are calculated to be $\lambda_{\text{eff}}$ = 505 nm and
$l_{\text{decay}}$ = 4960 nm.
In order to simulate far-field images of the wire near-field intensities are
recorded 10 nm below the structure and projected into the far-field
\cite{taflove_computational_2005, novotny_principles_2012}.
Simulated and experimentally obtained far-field images for a wire of 8 $\mu$m
length are displayed in Fig.~\ref{fig:measurement} (a) and (b), respectively.
The effect of the beam block used for spatially blocking the high-intensity
reflection spot resulting from the focused laser excitation of the nanowire
input terminal is visible by the nearly circular areas of reduced intensity
around the excitation spot.
Apart from a somewhat increased scattering in the experiment simulated and
experimental images agree exceptionally well.

For a detailed analysis of plasmon transmission through the nanowires we
extract for each nanowire the simulated far-field intensity as well as
experimental CCD image counts by integrating 1 $ \mu $m x 1 $ \mu $m squares
centered at the wire end (red and blue squares in Fig.~\ref{fig:measurement}).
From all 306 measured nanowires, only about 1\% showed unexpected signals that
we attribute to structural defects (see SI) and therefore excluded them from
the further analysis. The resulting values for the experimentally determined
wire transmission
are plotted as a function of the wire length (Fig.~\ref{fig:impedance} (blue
``x'')).  A corresponding plot resulting from simulated far-field images can be found in the
SI. 
The data is normalized such that the resulting decay curve (interleaved dark
green line) passes an intensity value of 1/e at a wire length matching
$l_\text{decay}$.  The detected intensity as a function of the wire length
exhibits an exponential decay modulated by an oscillatory behavior.  A
short-wavelength and a superimposed longer-wavelength oscillation are
distinguishable.  We compare the experimental data to the proposed model by
fitting Eq.~\ref{eqn:intensity} to the data.
In order to reduce the number of free parameters the mode's
propagation properties ($\lambda_\text{eff}$ = 505~nm, $l_\text{decay}$ =
4960~nm) and its reflection coefficient at the wire termination ($R$ = 0.42,
$\phi = 1.39$) are obtained from FDFD and FDTD simulation,
respectively. The initial phase offsets of both air-wave and glass-wave with
respect to the propagating plasmon are set to a fixed value of $\pi$.
With these constraints the amplitudes of the launched wire plasmon, the
air-wave, and the glass-wave at the incoupling position remain the only free
parameters of the model. The resulting fit to the simulation data (see SI)
shows perfect agreement. The amplitude ratio of the three contributions is
adopted from this fit, so that the final fit to the experimental data has only
one remaining amplitude parameter. An additional length offset of 85 nm has to
be introduced, which accounts for a systematic difference between the nominal
and the actual wire length which is also observed in high-resolution SEM
measurements. 
The model is plotted as a red line in Fig.~\ref{fig:impedance}.
The Fourier transformation of the length-dependent data (inset of
Fig.~\ref{fig:impedance}) shows two distinct peaks.  The highest frequency
component corresponds to a periodicity of 252~nm and therefore matches $\lambda_\text{eff} / 2$ as anticipated by
the Fabry-P\'{e}rot model. The slower oscillation corresponding to a wavelength
of about 1300~nm corresponds to the calculated beating wavelength of the excitation's vacuum
wavelength and the Fabry-P\'{e}rot modulated plasmon emission at
$\lambda_\text{beating}^\text{air--fpt}$ = 1372~nm. We thus attribute it to the
interference of the spherical air-wave scattered at the wire end and the emitted plasmons. This origin is
further supported by numerical simulations using a mode source to directly
excite the plasmons without launching spherical waves and the resulting absence
of the beating (see SI). 
In addition, the model predicts a beating between the plasmon emission and the
glass-wave showing a periodicity of about
$\lambda_\text{beating}^\text{glass-fpt}$ = 6520~nm. This beating wavelength is
about the same as the length-difference between longest and shortest measured
wire ($\Delta L_\text{max}$ = 6100~nm) and close to the length scale of the
plasmonic intensity decay ($l_\text{decay}$ = 4960~nm).  While it is not
clearly resolved in the Fourier transformation data it is important to note
that - within the observation window - it will appear as an additional slope of
the exponential decay curve. Neglecting this additional component would -
depending on it's relative phase - lead to either an under- or overestimation
of the decay length.
This can best be seen in Fig.~\ref{fig:impedance} by the dashed light green
curve which oscillates (visible beating
$\lambda_\text{beating}^\text{air--fpt}$ = 1372~nm)
above the interleaved dark green decay curve because of the additional
intensity caused by the beating between the spherical wave in glass and the
Fabry-P\'{e}rot modulated plasmon emission.
To observe and distinguish these different contributions in the experiment the
requirements on the sample's geometrical precision are demanding.  By
validating our experimental data against a model that includes artificial
errors, i.e.~by comparing the residuals, we determine the upper limits for the
uncertainties (standard deviation) of our structures geometrical parameters and
experimental conditions, i.e.~wire length,
width as well as random intensity
fluctuations to be 32~nm, 8~nm and 7\%, respectively (see SI).
For the length of the wires this corresponds to a relative error of below 0.4\%, which is,
taking into account the non-conductive substrate, smaller than the
experimentally accessible resolution limit of current state of the art
SEM techniques.

%
%
We conclude that the high precision and reproducibility of the fabricated
nanowires allowed us to reveal the nonnegligible influence of air and substrate
waves on their apparent length-dependent transmission. Our experiments also
show that by inclusion of these additional waves simulated and experimentally
data agree quantitatively within the remaining small experimental
uncertainties. The effect of the air wave becomes apparent as a beating
superimposed to the Fabry-P\'{e}rot standing wave pattern. The role of the
substrate wave is less obvious. We show that the very long beating wavelength
of the substrate wave with the wire plasmon causes unavoidable uncertainty
for the fitting of the decay length because the corresponding oscillatory
behaviour cannot be captured even for longer wires since the overall damping
of the plasmon becomes too strong. The resulting uncertainty regarding the
starting phase of the substrate wave is likely responsible for measurements of
decay lengths that reported too long or too short decay
lengths\cite{kusar_measurement_2012, wild_propagation_2012}.

\bibliography{impedance}
\end{document}